\documentclass[dvipdfmx]{article}
\usepackage[dvipdfmx]{color}
\usepackage[dvipdfmx]{graphicx}
\usepackage{spconf,amsmath}
\usepackage{amssymb}
\usepackage{url}

\newcommand{\vm}[1]{\textit{\boldmath $#1$}}
\newcommand{\x}{\vm{x}}
\newcommand{\y}{\vm{y}}
\newcommand{\z}{\vm{z}}

\newcommand{\red}[1]{\textcolor{black}{#1}}
\newcommand{\blue}[1]{\textcolor{black}{#1}}
\newcommand{\reva}[1]{\blue{#1}}
\newcommand{\revm}[1]{\blue{#1}}
\newcommand{\revb}[1]{\red{#1}}

\title{WaveCycleGAN: SYNTHETIC-TO-NATURAL SPEECH WAVEFORM CONVERSION USING CYCLE-CONSISTENT ADVERSARIAL NETWORKS}
\name{Kou Tanaka, Takuhiro Kaneko, Nobukatsu Hojo, and Hirokazu Kameoka}
\address{NTT Communication Science Laboratories, NTT Corporation, Japan \\
Email: \{tanaka.ko, kaneko.takuhiro, hojo.nobukatsu, kameoka.hirokazu\}@lab.ntt.co.jp}
\begin{document}
%
\maketitle
\begin{abstract}
We propose a learning-based filter \reva{that allows us} to directly modify \revm{a synthetic speech waveform into a natural speech waveform.}
Speech-processing systems using a \reva{vocoder} framework such as statistical parametric speech synthesis and voice conversion are convenient especially for a limited \revm{number of} data because it is possible to represent and process interpretable acoustic features over a compact space, such as the fundamental frequency ($F_0$) and mel-cepstrum.
However, \revm{a} well-known problem \revm{that leads to} the quality degradation of generated speech is an over-smoothing effect \revm{that eliminates} some detailed structure of generated/converted acoustic features.
To address this issue, we propose a synthetic-to-natural speech waveform conversion \revm{technique that uses} cycle-consistent adversarial networks \revm{and} which does not require any explicit assumption about speech waveform in adversarial learning.
In contrast to current techniques, since our modification is performed at the waveform level, we expect that the proposed method \revm{will} also \revm{make it possible} to generate ``\revm{vocoder-less}'' sounding speech even if the input speech is synthesized \revm{using a} \reva{vocoder} framework.
The experimental results demonstrate that our proposed method \revm{can} 1) alleviate the over-smoothing effect of \reva{the} acoustic features \revm{despite the direct modification method used for the waveform} and 2) \reva{greatly} improve the naturalness of the generated speech sounds.
\end{abstract}
\section{Introduction}

Speech processing systems such as statistical parametric speech synthesis~\cite{zen2013statistical} and statistical voice conversion~\cite{toda2007voice} are well-known frameworks.
These approaches using \revm{a} \reva{vocoder} framework have a significant advantage, especially for a limited \revm{number of} data, because it is possible to represent interpretable acoustic features over a compact space, such as the fundamental frequency ($F_0$) and mel-cepstrum, which are lower dimensional acoustic features than \revm{a} short-term Fourier transform (STFT) spectrogram.
Although these systems aim to produce speech with \revm{a} quality indistinguishable from \revm{that of} clean and real speech, processed and synthesized speech \revm{can} usually \revm{be distinguished} from natural speech.
The realization of synthetic-to-natural speech waveform conversion provides significant benefit \revm{with} many speech processing \revm{approaches,} especially \revm{when} using \revm{a} \reva{vocoder} framework.
Three major factors reported in~\cite{zen2009statistical} degrade the speech synthesized by \revm{a} statistical parametric speech synthesis \revm{technique}: \revm{the} accuracy of acoustic models, over-smoothing, \revm{which eliminates} some detailed structure of generated/converted acoustic features, and vocoding.
In this paper, we focus on vocoding and over-smoothing.

\begin{figure}[t]
    \centering
    \includegraphics[width=90mm,clip]{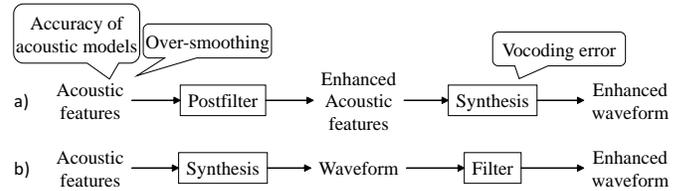}
    \caption{Three major factors~\cite{zen2009statistical} that degrade the quality of synthesized speech \revm{during} statistical parametric speech synthesis and general approaches to \revm{generating} more natural sounding speech by using \revm{post-processing}. Our proposed framework is assigned to a process b) which \revm{can} address not only the over-smoothing problem but also the vocoding error.}
    \label{fig:overview}
\end{figure}

To address the over-smoothing effect, several techniques \revm{for restoring the} fine structure of natural speech over acoustic features have been proposed~\cite{toda2007voice, takamichi2014postfilter, kaneko2017generative}.
These approaches, as shown in Fig.~\ref{fig:overview} a), have achieved significant improvements \revm{as regards} the naturalness of synthesized speech in the respective directions.
\revm{However, heuristics approaches such as enhancement of global variance~\cite{toda2007voice} and modulation spectrum~\cite{takamichi2014postfilter} are unsuitable for covering all the negative factors.}
On the other hand, although a learning-based postfilter~\cite{kaneko2017generative} enables \revm{us} to restore not only \revm{the} global variance and modulation spectrum but also other factors \revm{that} degrade the quality of synthesized speech, it is still \revm{insufficient} to generate natural speech because \revm{of the post filter needed not for the waveform but for the heuristic} acoustic features such as mel-cepstrum.
Furthermore, all of these approaches suffer from vocoding error because of the use of the \reva{vocoder} framework to synthesize the speech waveform.

To avoid this limitation, an end-to-end speech enhancement~\cite{pascual2017segan} method has been proposed within \revm{a} generative adversarial framework.
As shown in Fig.~\ref{fig:overview} b), since the waveform of the input speech was directly operated to obtain that of the desired speech after the vocoding part, \cite{pascual2017segan} \reva{has the potential to address not only the over-smoothing effect but also the vocoding error}.
Furthermore, \reva{the generative adversarial framework \revm{does not require us} to design any hand-crafted feature \revm{that creates} a gap between natural speech and synthetic speech, in advance.}
In preliminary experiments, we found that this method is \revm{unsuitable when the} alignments \footnote{\revm{In this paper, we define alignment considering both the magnitude information and the phase information of speech because we focus on modifying the speech waveform rather than the acoustic features.}} between the input waveform and the desired waveform are not perfect.
For example, \revm{the} noise reduction \revm{of} noisy speech simulated by adding noise to the speech waveform recorded in \revm{an} ideal environment \revm{succeeded} because of the perfect alignment between the simulated noisy speech and the clean source speech.
However, \revm{the} conversion \revm{of} synthetic speech generated by text-to-speech synthesis and voice conversion processing to natural speech is not easy to \revm{achieve} by applying this method \revm{because of} the alignment problem as mentioned above.

In this paper, we propose a learning-based filter \reva{that allows us} to convert \revm{a} synthetic speech waveform into \revm{a} natural speech waveform using cycle-consistent adversarial networks with a fully convolutional architecture. \revm{We} adopt cycle-consistent adversarial networks \revm{because they do not} require a dataset forcibly paired at the time frame level and as the name implies, \revm{they} are trained within the adversarial learning.
\revb{In contrast to ~\cite{kaneko2017parallel} which is also inspired by the cycle-consistent adversarial networks~\cite{zhu2017unpaired} to convert not speech waveform but acoustic feature, since our modification is performed at the waveform level, we expect that the proposed method will make it possible to generate ``vocoder-less'' sounding speech even if the input speech is synthesized using a vocoder framework.}
Furthermore, we adopt a gated convolutional neural network (CNN) architecture~\cite{dauphin2016language}, which is able to capture long- and short-term dependencies in the speech waveform.
The experimental results demonstrate that our proposed method \revm{can} 1) alleviate the over-smoothing effect of \reva{the} acoustic features \revm{despite the direct modification method used for the waveform} and 2) \reva{greatly} improve the naturalness of the generated speech sounds.

\section{SEGAN: Speech Enhancement Generative Adversarial Network}

\subsection{Generative Adversarial Networks} 
Generative Adversarial Networks (GANs)~\cite{goodfellow2014generative} are generative models consisting of two neural networks.
One is a generator $G$ that \revm{learns} to convert a sample $\z$ from a prior distribution $P(\z)$ to a target sample $\x$ from a distribution $P_{\mathrm{Data}(\x)}$, which is a sample \revm{from} the training data.
The generator aims to learn a projection that \revm{can} imitate the true feature distribution and to generate samples related to the training data.
\revm{The other} is a discriminator $D$ that \revm{learns the} boundary between imitated features generated by the generator $G$ and true features picked up from the training data.

The adversarial characteristic \revm{arises} from the fact that the discriminator $D$ tries to classify the instances $\x$ \revm{obtained} from the true data distribution $P_\mathrm{Data}(\x)$ as real and the candidates $G(\z)$ produced by the generator $G$ as fake, while the generator $G$ tries to make the discriminator $D$ classify those $G(\z)$ as real.
Through back-propagation, the generator $G$ becomes able to generate better candidates $G(\z)$ and the discriminator $D$ becomes able to distinguish the generated ones $G(\z)$ and real data $\x$.
The objective function of the adversarial learning is formulated as the following minimax game between $G$ and $D$,
\begin{align}
    \min_G \max_D {\mathcal L}_\mathrm{gan} & (G_{\z \to \x}, D_\x) \nonumber \\
    & = \mathbb{E}_{x \sim P_{\mathrm{Data}(x)}} [\log D_\x (\x)] \nonumber \\
    & + \mathbb{E}_{z \sim P_{z}(z)} [\log (1 - D_\x (G_{\z \to \x}(\z)))]. \label{eq:gan}
\end{align}

\reva{Although the GANs achieve state-of-the-art results \revm{in} a variety of generative tasks~\cite{radford2015unsupervised,saito2018statistical},} the difficulty of the training is a well-known problem.
For instance, the classic approach suffers from a vanishing gradient problem due to the sigmoid cross-entropy loss used for training.
\revm{Several adversarial training techniques have been proposed to overcome this difficulty.}
The least-squares GAN (LSGAN) approach~\cite{mao2017least} stabilizes the training process by replacing the cross-entropy loss shown in Eq.~\ref{eq:gan} with the least-squares function as follows.
\begin{align}
    \min_D {\mathcal L}_\mathrm{lsgan} (D_\x) 
    & = \frac{1}{2} \mathbb{E}_{x \sim P_{\mathrm{Data}(x)}} [(D_\x(\x)-1)^2] \nonumber \\
    & + \frac{1}{2} \mathbb{E}_{z \sim P_{z}(z)} [D_\x (G_{\z \to \x}(\z))^2], \label{eq:lsgan_d} \\
    \min_G {\mathcal L}_\mathrm{lsgan} (G_{\z \to \x})
    & = \frac{1}{2} \mathbb{E}_{z \sim P_{z}(z)} [(D_\x (G_{\z \to \x}(\z))-1)^2]. \label{eq:lsgan_g}
\end{align}

\subsection{GANs for Speech Enhancement} 

\begin{figure}[t]
    \centering
    \includegraphics[width=90mm,clip]{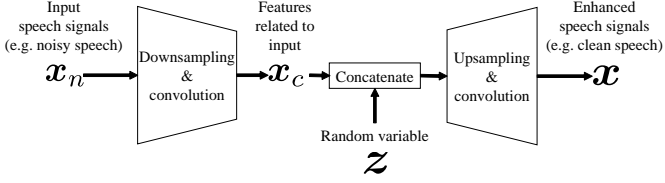}
    \caption{Generator network for speech enhancement reported in~\cite{pascual2017segan}. Structure is similar to an auto-encoder.}
    \label{fig:segan}
\end{figure}

To \revm{retain} the linguistic information of speech samples, ~\cite{pascual2017segan} adopts a conditioned version GAN \revm{that} has some extra information in $G$ and $D$ to perform mapping and classification.
As shown in Fig.~\ref{fig:segan}, in the structure of the generator $G$, which is similar to an auto-encoder, \revm{a} noisy speech signal $\x_\mathrm{n}$, which is the input of the $G$ network, \revm{is} encoded \revm{as} $\x_c$.
After concatenating the random vector $\z$ with the encoded \revm{vector} $\x_c$, \revm{which is} treated as \revm{a} conditional vector, the decoding part of \revm{the} $G$ network is performed as transposed convolutions (a.k.a. deconvolutions or fractionally strided convolutions) to obtain the enhanced \revm{vector}.

To achieve the generation of speech samples \revm{that are} closer to clean speech, a secondary component is added to the loss of $G$.
~\cite{pascual2017segan} adopts the L1 norm, as it has been proven to be effective in the image manipulation domain ~\cite{isola2017image, pathak2016context}.
\revm{In this way}, they \revm{allow} the adversarial component to add more fine-grained and realistic results.
A new hyper-parameter $\lambda_\mathrm{SEGAN}$ controls the magnitude of the L1 norm.
Finally, the loss function of the generator $G$ becomes
\begin{align}
    & \min_G {\mathcal L}_\mathrm{lsgan} (G_{\x_\mathrm{n},\z \to \x}) \nonumber \\
    & = \frac{1}{2} \mathbb{E}_{z \sim P_{z}(z), x_\mathrm{n} \sim P_{Data(x_\mathrm{n})}} [(D_\x (G_{\x_\mathrm{n},\z \to \x}(\x_\mathrm{n},\z))-1)^2] \nonumber \\
    & + \lambda_\mathrm{SEGAN}||G_{\x_\mathrm{n},\z \to \x}(\x_\mathrm{n},\z)-\x||_1. \label{eq:segan_g}
\end{align}

\section{Synthetic-to-Natural Speech Waveform Conversion Using Cycle-Consistent Adversarial Networks}

\subsection{Concept} 
In preliminary experiments, we found that SEGAN~\cite{pascual2017segan} could not be easily applied to the conversion \revm{of a} synthetic speech waveform to \revm{a} natural speech waveform.
\reva{One possible reason is that the misalignment caused by the different lengths and generation processes of synthetic and natural speech makes it difficult to ensure the \revm{operation of the} bijective function in the generator $G$.
\revm{Specifically}, the phase information of the speech waveform synthesized by using the vocoder framework is \revm{very} far from that of natural speech, even if the magnitude information of the synthetic speech is close to that of natural speech.
We assume that these factors \revm{induce} ``mode collapse'', which is a well-known problem \revm{when} training GANs, and the SEGAN does not guarantee that an individual input and output are paired up in a meaningful way.
\revm{Generally} speaking, all input speech signals map to the same output speech signals and the optimization fails to make progress~\cite{goodfellow2014generative}.}

To solve this problem, we focus on cycle-consistent adversarial networks~\cite{zhu2017unpaired}.
This approach has introduced \revm{a} ``cycle consistent'' \revm{property,} which \revm{ensures return to} the original \revm{sample}~\cite{brislin1970back}.
Mathematically, if we have a converter $G_{\x \to \y}$ and another converter $G_{\y \to \x}$, $G_{\x \to \y}$ and $G_{\y \to \x}$ should be \revm{the} inverse of each other, and both mappings should be bijections.
We incorporate this property into SEGAN by training the mapping functions $G_{\x \to \y}$ and $G_{\y \to \x}$ simultaneously and adding a cycle consistency loss~\cite{zhou2016learning} that encourages $G_{\y \to \x}(G_{\x \to \y}(\x)) \approx \x$ and $G_{\x \to \y}(G_{\y \to \x}(\y)) \approx \y$.
Combining the cycle consistency loss with the adversarial losses defines our full objective for a training procedure using perfect alignment.

Furthermore, we focus on a convolutional architecture called \revm{a} gated CNN.
The gated CNN has recently been shown to be powerful \revm{for} modeling long-term sequential data.
It was originally introduced for language modeling and was shown to outperform long short-term memory (LSTM) language models trained in a similar setting~\cite{dauphin2016language}.
We previously applied a gated CNN architecture for acoustic feature sequence modeling, and its effectiveness has already been confirmed~\cite{kaneko2017sequence, tanaka2018vae-space}.
With a gated CNN, the output of a hidden layer of a network is described as a linear projection modulated by an output gate.
Similar to an LSTM~\cite{hochreiter1997long} and \revm{a} gated recurrent unit (GRU)~\cite{cho2014learning}, the output gate controls what information should be propagated through the hierarchy of layers and allows \revm{the capture of} long-term structures.

\subsection{Cycle-Consistent Adversarial Networks} 

\begin{figure}[t]
    \centering
    \includegraphics[width=80mm]{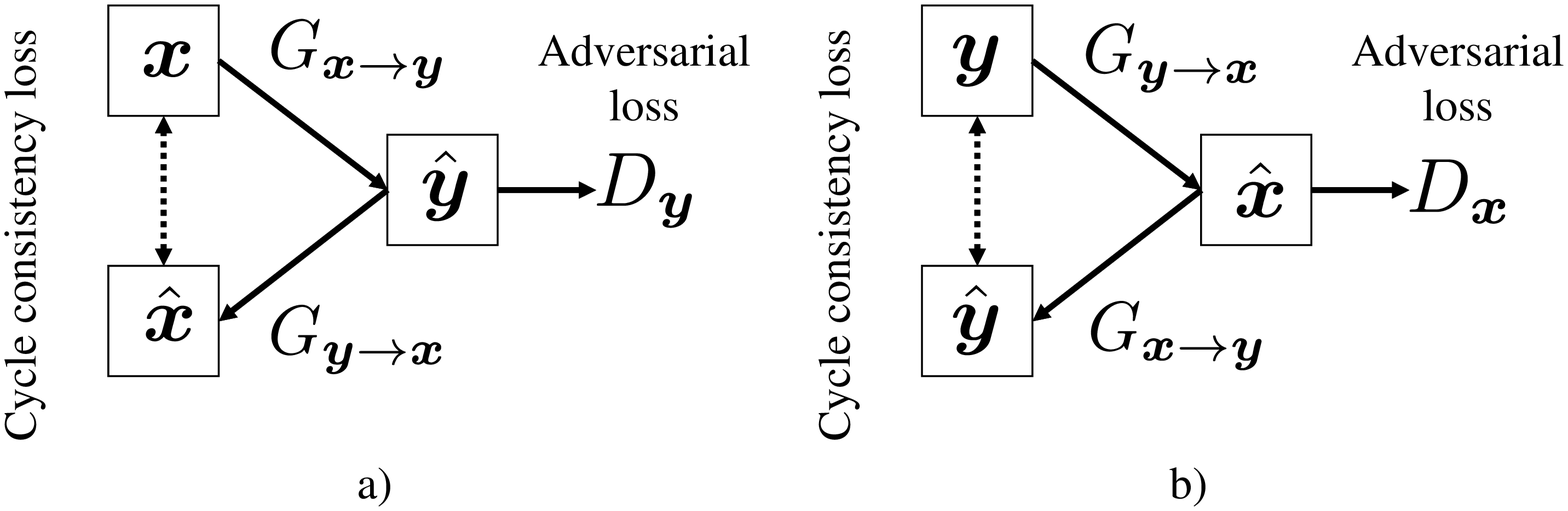}
    \caption{Training procedures of cycle-consistent adversarial networks: a) Forward-inverse mapping to consider forward cycle consistency and b) inverse-forward mapping to consider backward cycle consistency.}
    \label{fig:cyclegan}
\end{figure}

For each speech sample $\x$, the speech waveform conversion cycle shown in Fig.~\ref{fig:cyclegan} a) constrains the samples $\x$ \revm{to return} to the original speech through a target domain corresponding to the samples $\y$, $\x \to G_{\x \to \y}(\x) \to G_{\y \to \x}(G_{\x \to \y}(\x)) \approx \x$.
This cycle consistency is called forward cycle consistency.
Similarly, as shown in Fig.~\ref{fig:cyclegan} b), for each speech waveform $\y$, $G_{\x \to \y}$ and $G_{\y \to \x}$ are constrained by a backward cycle consistency, $\y \to G_{\y \to \x}(\y) \to G_{\x \to \y}(G_{\y \to \x}(\y)) \approx \y$.
Therefore, these are described as the following cycle consistency loss,
\begin{align}
    {\mathcal L}_\mathrm{cyc} & = \mathbb{E}_{x \sim P_{\mathrm{Data}(x)}} [|| G_{\y \to \x}(G_{\x \to \y}(\x)) - \x ||_1] \nonumber \\
    & + \mathbb{E}_{y \sim P_{\mathrm{Data}(y)}} [|| G_{\x \to \y}(G_{\y \to \x}(\y)) - \y ||_1].
\end{align}

\noindent Finally, the objective function is
\begin{align}
    {\mathcal L}_\mathrm{full} & = {\mathcal L}_\mathrm{gan}(G_{\x \to \y}, D_\y) + {\mathcal L}_\mathrm{gan}(G_{\y \to \x}, D_\x) \nonumber \\
    & + \lambda_\mathrm{cyc} {\mathcal L}_\mathrm{cyc}, \label{eq:cyclegan}
\end{align}
where $\lambda_\mathrm{cyc}$ is a hyper parameter used to control the cycle consistency loss.

\begin{figure*}[t]
    \centering
    \includegraphics[width=180mm,clip]{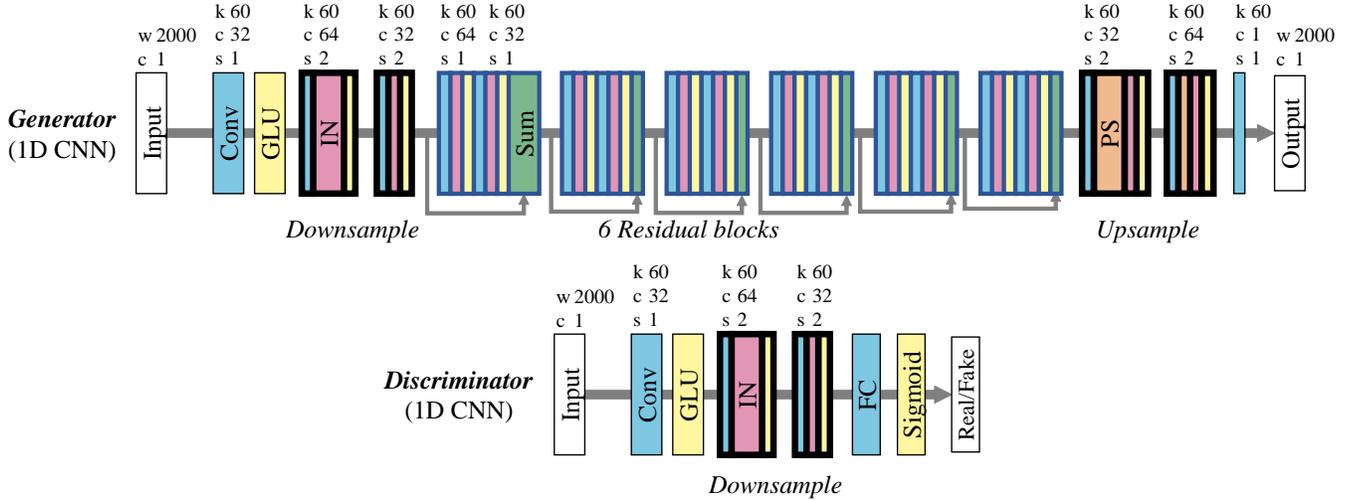}
    \caption{Network architectures of generator G and discriminator D. ``Conv'', ``GLU'', ``IN'', ``PS'', ``FC'', and ``Sigmoid'' denote convolution, instance normalization, gated linear unit, pixel shuffler, fully connected, and sigmoid layers, respectively. In an input or output layer, w and c represent width and number of channels, respectively. In each convolutional layer, k, c, and s denote kernel size, number of channels, and stride size, respectively.}
    \label{fig:architecture}
\end{figure*}

\subsection{Identity-Mapping Loss} 

Cycle consistency loss \revm{allows us} to reduce the possible mapping functions by constraining a structure.
However, in \revm{a} waveform modification task, the linguistic information is not always preserved by incorporating only the cycle consistency loss.
The identity-mapping loss reported in~\cite{taigman2016unsupervised} preserves \revm{the} compositions \revm{of} the input samples and the converted samples.
~\cite{zhu2017unpaired} has applied \revm{this approach} to color preservation and demonstrated its effectiveness.
\reva{Note that the secondary component of Eq. \ref{eq:segan_g} is also identity-mapping loss.}
\revm{To encourage the generators $G_{\x \to \y}$ and $G_{\y \to \x}$ to preserve linguistic information,} we also incorporate this property as follows.
\begin{align}
    {\mathcal L}_\mathrm{id} & = \mathbb{E}_{x \sim P_{\mathrm{Data}(x)}} [|| G_{\y \to \x}(\x) - \x ||_1] \nonumber \\
    & + \mathbb{E}_{y \sim P_{\mathrm{Data}(y)}} [|| G_{\x \to \y}(\y) - \y ||_1].
\end{align}

\noindent In practice, the weighted loss $\lambda_\mathrm{id}{\mathcal L}_\mathrm{id}$ with a hyper parameter $\lambda_\mathrm{id}$ to control the identity-mapping loss is added to Eq.~\ref{eq:cyclegan}.

\subsection{Sequential Modeling with Gated CNN}

To capture long- and short-term dependencies in speech waveforms, we use a gated CNN~\cite{dauphin2016language} to construct both the generator and discriminator networks of the GAN.
The gated CNNs are CNNs equipped with gated linear units (GLUs) as activation functions instead of the regular rectified linear units (ReLUs)~\cite{nair2010rectified} or Tanh activations.
The output of the $l_\mathrm{th}$ hidden layer of a gated CNN is described as a linear projection $\vm{H}_{l-1} \ast \vm{W}_l+\vm{b}_l$ modulated by an output gate $\sigma(\vm{H}_{l-1} \ast \vm{V}_l + \vm{c}_l)$
\begin{align}
    \vm{H}_l = (\vm{H}_{l-1} \ast \vm{W}_l + \vm{b}_l) \otimes \sigma (\vm{H}_{l-1} \ast \vm{V}_l + \vm{c}_l),
\end{align}

\noindent where $\vm{W}_l$, $\vm{V}_l$, $\vm{b}_l$ and $\vm{c}_l$ are the network parameters to be trained, $\sigma$ is the sigmoid function and $\otimes$ indicates the element-wise product.
Similar to LSTMs, the output gate multiplies each element of $\vm{H}_{l-1} \ast \vm{W}_l + \vm{b}_l$ and controls what information should be propagated through the hierarchy of layers in a data-driven manner.

\section{Experimental Evaluation} 

\subsection{Experimental Conditions} 

\noindent {\bf Datasets (Natural):}
We used \revm{a} Japanese speech dataset \revm{consisting of utterances by} one professional female narrator.
To evaluate the performance, we used 30 sentences (speech sections of 5.3 minutes).
To train the models, we used about 6,500 sentences for a baseline system and 400 sentences (speech sections of 1.2 hours) for the conventional and proposed methods.
The sampling rate of the speech signals was 22.05 kHz.
Audio samples \revm{can be accessed on} our web page\footnote{\url{http://www.kecl.ntt.co.jp/people/tanaka.ko/projects/s2n/s2n_speech_waveform_conversion.html}}.

\noindent {\bf Baseline system (Baseline):}
We used a DNN-based statistical parametric speech synthesis method~\cite{zen2013statistical} as the baseline.
From the speech data, 40 Mel-cepstral coefficients, logarithmic $F_0$, and 5-band aperiodicities were extracted every 5 ms \revm{with} the STRAIGHT analysis system~\cite{kawahara1999restructuring, kawahara2001aperiodicity}.
\revm{The contextual features used} as the input were 506-dimensional linguistic features including phonemes and mora positions.
The output \revm{consisted of} 40 Mel-cepstral coefficients, log $F_0$, 5-band aperiodicities, their delta and delta-delta features, and a voiced/unvoiced binary value.
The DNN architectures were feed-forward networks including 5 hidden layers \revm{each with} 1,024 units.

\begin{figure}[t]
    \centering
    \includegraphics[trim=0cm 0cm 0cm 0.7cm, width=80mm,clip]{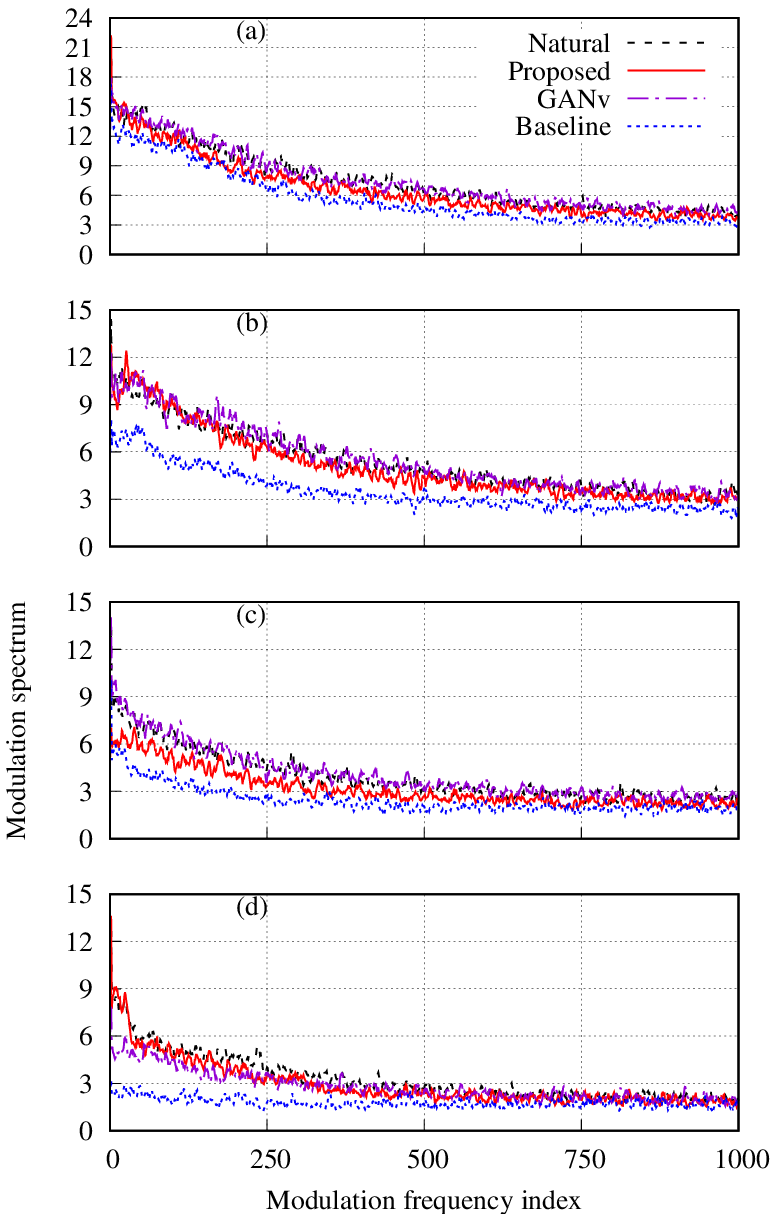}
    \caption{Average modulation spectrums of the first 1k indices for a) 10th, b) 20th, c) 30th and d) 40th mel-cepstral coefficient sequences.}
    \label{fig:ms14}
\end{figure}

\noindent {\bf Conventional method (GANv):}
\reva{As a conventional approach, we used a GAN-based postfilter ~\cite{kaneko2017generative} \revm{not for} the speech waveform \revm{but for} the acoustic features.
The system setting was the same as the reported setting, except for the excitation signals.
Although ~\cite{kaneko2017generative} used the excitation signals of natural speech, we used the excitation signals generated by the vocoding for evaluating \revm{all of the} synthetic speech.
We applied the conventional method \revm{only to} voiced segments.
}

\noindent {\bf Our proposed method (Proposed):}
We designed a network based on recent success \revm{of} image modeling~\cite{ledig2016photo}.
Figure~\ref{fig:architecture} shows the network architectures of our proposed model.
The network included downsampling layers, residual blocks~\cite{he2016deep}, and upsampling layers.
We used instance normalization (IN)~\cite{dmitry2016instance}, instead of batch normalization~\cite{ioffe2015batch}.
We used pixel shuffler (PS) for upsampling \revm{where} the effectiveness was demonstrated in high-resolution image generation~\cite{ledig2016photo}.
We normalized the speech waveform to \revm{zero mean and unit variance} using their training sets.
To stabilize \revm{the} training, we used a least squares GAN~\cite{mao2017least}.
We set $\lambda_\mathrm{cyc}$ \revm{at} 10.
To guide the learning process, we set $\lambda_\mathrm{id}$ \revm{at} 5 for the first 20k iterations and \revm{linearly decay to} 0 over the next 20k iterations.
We optimized the model parameters using the Adam optimizer~\cite{kingma2014adam} with a mini-batch of size 32.
The learning parameters $\alpha$ were set \revm{at} 0.0001 for discriminators and 0.0002 for generators.
We used the same learning rate for the first 250k iterations and linearly decay to 0 over the next 250k iterations.
The other learning parameters of the Adam optimizer, $\beta_1$ and $\beta_2$, were set \revm{at} 0.5 and 0.99, respectively.
Note that since the generators are fully convolutional, \revm{they can handle an arbitrary length input}.

\subsection{Modulation Spectrum over Acoustic Features} 
\label{sec:eval_ms}

To confirm the alleviation of the over-smoothing effect of the acoustic features, we \revm{applied} the conventional and proposed methods to speech synthesized by the baseline system and \revm{obtained} modulation spectrums of mel-cepstrum sequences on each system.
\revm{Although} the modulation spectrum is traditionally defined as a value calculated using the Fourier transform of the parameter sequence~\cite{Atlas2003}, this paper defines the modulation spectrum as its \reva{logarithmic power spectrum.}
\reva{We used 8,192 FFT points.}

The average modulation spectrums \reva{of the first 1k indices for} the 10th, 20th, 30th and 40th mel-cepstral coefficient sequences \revm{are} shown in Fig.~\ref{fig:ms14}.
We found that {\bf Baseline} suffered \revm{more} from the over-smoothing effect \revm{than} {\bf GANv} and {\bf Natural}.
On the other hand, {\bf GANv} and {\bf Proposed} are close to {\bf Natural}.
\reva{\revm{As with} the GAN-based postfilter for the acoustic feature {\bf GANv}, the result demonstrated that our proposed method for the speech waveform {\bf Proposed} successfully alleviated the over-smoothing effect caused by the statistical parametric speech synthesis process.}

\subsection{Subjective Evaluation for Naturalness} 

\begin{figure}[t]
    \centering
    \includegraphics[width=80mm,clip]{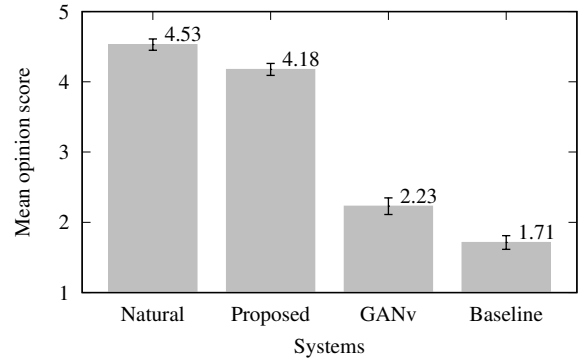}
    \caption{Subjective 5-scale mean opinion score \revm{regarding} naturalness, with 95\% confidence intervals.}
    \label{fig:eval_s_naturalness}
\end{figure}

We conducted a subjective 5-scale mean opinion score test \revm{regarding} the naturalness of the generated speech.
\reva{10} listeners participated and each listener evaluated \reva{120 speech samples (30 speech samples $\times$ 4 systems).}
We applied the conventional and proposed methods to the same speech waveform {\bf Baseline} as in Sec.~\ref{sec:eval_ms}.

Figure~\ref{fig:eval_s_naturalness} shows that our proposed method {\bf Proposed} achieved a significant improvement \revm{in terms} of the naturalness of the generated speech, compared with {\bf Baseline} and {\bf GANv}.
\reva{This result indicates that our approach is more effective \revm{than} the \revm{use of} postfilters for the acoustic features because it is possible to address both the over-smoothing problem and the vocoding error.}
Furthermore, \revm{with} {\bf Proposed}, the listeners commented that the ``buzzy'' sound peculiar to vocoding was sufficiently improved.
However, there is still a gap between {\bf Proposed} and {\bf Natural}.
\reva{One possible reason \revm{for} the gap is the “hoarse” sound of {\bf Proposed}.
The listeners also advised that {\bf Proposed} was distinguishable from {\bf Natural} because {\bf Proposed} sometimes \revm{had a}  ``hoarse'' sound.}

\section{Conclusion} 
In this paper, to realize a synthetic-to-natural speech filter, we proposed a learning-based filter \reva{that allows us} to convert a synthetic speech waveform into a natural speech waveform using cycle-consistent adversarial networks.
Since our process was \revm{applied} after the synthesis part in statistical parametric speech synthesis, we expected that our approach \revm{would be able} to address not only the over-smoothing problem but also the vocoding error.
The experimental results demonstrated that our proposed method 1) alleviated the over-smoothing effect of the acoustic features \revm{despite} the direct modification method used for the waveform and 2) \reva{dramatically} improved the naturalness of the generated speech sounds.
In the future, we will further fill the gap between natural speech and synthetic speech by considering the auditory property.

\section*{Acknowledgment} 
This work was supported by JSPS KAKENHI 17H01763.

\bibliographystyle{IEEEbib}

\begin{thebibliography}{10}

\bibitem{zen2013statistical}
Heiga Zen, Andrew Senior, and Mike Schuster,
\newblock ``Statistical parametric speech synthesis using deep neural
  networks,''
\newblock in {\em Acoustics, Speech and Signal Processing (ICASSP), 2013 IEEE
  International Conference on}. IEEE, 2013, pp. 7962--7966.

\bibitem{toda2007voice}
Tomoki Toda, Alan~W Black, and Keiichi Tokuda,
\newblock ``Voice conversion based on maximum-likelihood estimation of spectral
  parameter trajectory,''
\newblock {\em IEEE Transactions on Audio, Speech, and Language Processing},
  vol. 15, no. 8, pp. 2222--2235, 2007.

\bibitem{zen2009statistical}
Heiga Zen, Keiichi Tokuda, and Alan~W Black,
\newblock ``Statistical parametric speech synthesis,''
\newblock {\em Speech Communication}, vol. 51, no. 11, pp. 1039--1064, 2009.

\bibitem{takamichi2014postfilter}
Shinnosuke Takamichi, Tomoki Toda, Graham Neubig, Sakriani Sakti, and Satoshi
  Nakamura,
\newblock ``A postfilter to modify the modulation spectrum in {HMM}-based
  speech synthesis,''
\newblock in {\em Acoustics, Speech and Signal Processing (ICASSP), 2014 IEEE
  International Conference on}. IEEE, 2014, pp. 290--294.

\bibitem{kaneko2017generative}
Takuhiro Kaneko, Hirokazu Kameoka, Nobukatsu Hojo, Yusuke Ijima, Kaoru
  Hiramatsu, and Kunio Kashino,
\newblock ``Generative adversarial network-based postfilter for statistical
  parametric speech synthesis,''
\newblock in {\em Proc. 2017 IEEE International Conference on Acoustics, Speech
  and Signal Processing (ICASSP2017)}, 2017, pp. 4910--4914.

\bibitem{pascual2017segan}
Santiago Pascual, Antonio Bonafonte, and Joan Serr{\`a},
\newblock ``{SEGAN}: Speech enhancement generative adversarial network,''
\newblock {\em arXiv preprint arXiv:1703.09452}, 2017.

\bibitem{kaneko2017parallel}
Takuhiro Kaneko and Hirokazu Kameoka,
\newblock ``Parallel-data-free voice conversion using cycle-consistent
  adversarial networks,''
\newblock {\em arXiv preprint arXiv:1711.11293}, 2017.

\bibitem{zhu2017unpaired}
Jun-Yan Zhu, Taesung Park, Phillip Isola, and Alexei~A Efros,
\newblock ``Unpaired image-to-image translation using cycle-consistent
  adversarial networks,''
\newblock {\em arXiv preprint arXiv:1703.10593}, 2017.

\bibitem{dauphin2016language}
Yann~N Dauphin, Angela Fan, Michael Auli, and David Grangier,
\newblock ``Language modeling with gated convolutional networks,''
\newblock {\em arXiv preprint arXiv:1612.08083}, 2016.

\bibitem{goodfellow2014generative}
Ian Goodfellow, Jean Pouget-Abadie, Mehdi Mirza, Bing Xu, David Warde-Farley,
  Sherjil Ozair, Aaron Courville, and Yoshua Bengio,
\newblock ``Generative adversarial nets,''
\newblock in {\em Advances in neural information processing systems}, 2014, pp.
  2672--2680.

\bibitem{radford2015unsupervised}
Alec Radford, Luke Metz, and Soumith Chintala,
\newblock ``Unsupervised representation learning with deep convolutional
  generative adversarial networks,''
\newblock {\em arXiv preprint arXiv:1511.06434}, 2015.

\bibitem{saito2018statistical}
Yuki Saito, Shinnosuke Takamichi, Hiroshi Saruwatari, Yuki Saito, Shinnosuke
  Takamichi, and Hiroshi Saruwatari,
\newblock ``Statistical parametric speech synthesis incorporating generative
  adversarial networks,''
\newblock {\em IEEE/ACM Transactions on Audio, Speech and Language Processing
  (TASLP)}, vol. 26, no. 1, pp. 84--96, 2018.

\bibitem{mao2017least}
Xudong Mao, Qing Li, Haoran Xie, Raymond~YK Lau, Zhen Wang, and Stephen~Paul
  Smolley,
\newblock ``Least squares generative adversarial networks,''
\newblock in {\em 2017 IEEE International Conference on Computer Vision
  (ICCV)}. IEEE, 2017, pp. 2813--2821.

\bibitem{isola2017image}
Phillip Isola, Jun-Yan Zhu, Tinghui Zhou, and Alexei~A Efros,
\newblock ``Image-to-image translation with conditional adversarial networks,''
\newblock {\em arXiv preprint}, 2017.

\bibitem{pathak2016context}
Deepak Pathak, Philipp Krahenbuhl, Jeff Donahue, Trevor Darrell, and Alexei~A
  Efros,
\newblock ``Context encoders: Feature learning by inpainting,''
\newblock in {\em Proceedings of the IEEE Conference on Computer Vision and
  Pattern Recognition}, 2016, pp. 2536--2544.

\bibitem{brislin1970back}
Richard~W Brislin,
\newblock ``Back-translation for cross-cultural research,''
\newblock {\em Journal of cross-cultural psychology}, vol. 1, no. 3, pp.
  185--216, 1970.

\bibitem{zhou2016learning}
Tinghui Zhou, Philipp Krahenbuhl, Mathieu Aubry, Qixing Huang, and Alexei~A
  Efros,
\newblock ``Learning dense correspondence via 3{D}-guided cycle consistency,''
\newblock in {\em Proceedings of the IEEE Conference on Computer Vision and
  Pattern Recognition}, 2016, pp. 117--126.

\bibitem{kaneko2017sequence}
Takuhiro Kaneko, Hirokazu Kameoka, Kaoru Hiramatsu, and Kunio Kashino,
\newblock ``Sequence-to-sequence voice conversion with similarity metric
  learned using generative adversarial networks,''
\newblock {\em Proc. Interspeech 2017}, pp. 1283--1287, 2017.

\bibitem{tanaka2018vae-space}
Kou Tanaka, Hirokazu Kameoka, and Kazuho Morikawa,
\newblock ``{VAE-SPACE}: Deep generative model of voice fundamental frequency
  contours,''
\newblock {\em Proc. ICASSP 2018}, 2018.

\bibitem{hochreiter1997long}
Sepp Hochreiter and J{\"u}rgen Schmidhuber,
\newblock ``Long short-term memory,''
\newblock {\em Neural computation}, vol. 9, no. 8, pp. 1735--1780, 1997.

\bibitem{cho2014learning}
Kyunghyun Cho, Bart Van~Merri{\"e}nboer, Caglar Gulcehre, Dzmitry Bahdanau,
  Fethi Bougares, Holger Schwenk, and Yoshua Bengio,
\newblock ``Learning phrase representations using {RNN} encoder-decoder for
  statistical machine translation,''
\newblock {\em arXiv preprint arXiv:1406.1078}, 2014.

\bibitem{taigman2016unsupervised}
Yaniv Taigman, Adam Polyak, and Lior Wolf,
\newblock ``Unsupervised cross-domain image generation,''
\newblock {\em arXiv preprint arXiv:1611.02200}, 2016.

\bibitem{nair2010rectified}
Vinod Nair and Geoffrey~E Hinton,
\newblock ``Rectified linear units improve restricted boltzmann machines,''
\newblock in {\em Proceedings of the 27th international conference on machine
  learning (ICML-10)}, 2010, pp. 807--814.

\bibitem{kawahara1999restructuring}
Hideki Kawahara, Ikuyo Masuda-Katsuse, and Alain De~Cheveigne,
\newblock ``Restructuring speech representations using a pitch-adaptive
  time--frequency smoothing and an instantaneous-frequency-based f0 extraction:
  Possible role of a repetitive structure in sounds,''
\newblock {\em Speech communication}, vol. 27, no. 3, pp. 187--207, 1999.

\bibitem{kawahara2001aperiodicity}
Hideki Kawahara, Jo~Estill, and Osamu Fujimura,
\newblock ``Aperiodicity extraction and control using mixed mode excitation and
  group delay manipulation for a high quality speech analysis, modification and
  synthesis system straight,''
\newblock in {\em Second International Workshop on Models and Analysis of Vocal
  Emissions for Biomedical Applications}, 2001.

\bibitem{ledig2016photo}
Christian Ledig, Lucas Theis, Ferenc Husz{\'a}r, Jose Caballero, Andrew
  Cunningham, Alejandro Acosta, Andrew Aitken, Alykhan Tejani, Johannes Totz,
  Zehan Wang, et~al.,
\newblock ``Photo-realistic single image super-resolution using a generative
  adversarial network,''
\newblock {\em arXiv preprint}, 2016.

\bibitem{he2016deep}
Kaiming He, Xiangyu Zhang, Shaoqing Ren, and Jian Sun,
\newblock ``Deep residual learning for image recognition,''
\newblock in {\em Proceedings of the IEEE conference on computer vision and
  pattern recognition}, 2016, pp. 770--778.

\bibitem{dmitry2016instance}
Dmitry Ulyanov, Andrea Vedaldi, and Victor~S. Lempitsky,
\newblock ``Instance normalization: The missing ingredient for fast
  stylization,''
\newblock {\em CoRR}, vol. abs/1607.08022, 2016.

\bibitem{ioffe2015batch}
Sergey Ioffe and Christian Szegedy,
\newblock ``Batch normalization: Accelerating deep network training by reducing
  internal covariate shift,''
\newblock in {\em International conference on machine learning}, 2015, pp.
  448--456.

\bibitem{kingma2014adam}
Diederik Kingma and Jimmy Ba,
\newblock ``Adam: A method for stochastic optimization,''
\newblock {\em arXiv preprint arXiv:1412.6980}, 2014.

\bibitem{Atlas2003}
Les Atlas and Shihab~A. Shamma,
\newblock ``Joint acoustic and modulation frequency,''
\newblock {\em EURASIP Journal on Advances in Signal Processing}, vol. 2003,
  no. 7, pp. 310290, June 2003.

\end{thebibliography}

\end{document}